\documentclass[3p,times,twocolumn]{elsarticle}

\usepackage{ecrc}

\volume{00}

\firstpage{1}

\journalname{Nuclear and Particle Physics Proceedings}

\runauth{}

\jid{nppp}

\jnltitlelogo{Nuclear and Particle Physics Proceedings}

\usepackage{amssymb}

\usepackage{lineno}

\usepackage[figuresright]{rotating}
\usepackage{xspace}

\newcommand*{\pp}{\ensuremath{pp}\xspace}

\newcommand*{\PbPb}{Pb+Pb\xspace}

\newcommand*{\AuAu}{Au+Au\xspace}

\newcommand*{\CuCu}{Cu+Cu\xspace}

\newcommand*{\pA}{\ensuremath{p}+A\xspace}

\newcommand*{\dAu}{\ensuremath{d}+Au\xspace}

\newcommand*{\pPb}{\ensuremath{p}+\rm{Pb}\xspace}

\newcommand*{\sqn}{\ensuremath{\sqrt{s_{_{\mathrm{NN}}}}}\xspace}
\newcommand*{\sqs}{\ensuremath{\sqrt{s}}\xspace}

\newcommand*{\pT}{\ensuremath{p_{\mathrm{T}}}\xspace}

\begin{document}

\begin{frontmatter}



\dochead{}

\title{Measurement of long-range particle correlations in small systems with the ATLAS detector}


\author{Alexander Milov for the ATLAS Collaboration}

\address{Department of Particle Physics and Astrophysics, Weizmann Institute of Science, 234 Herzl str. Rehovot 7610001, Israel}

\begin{abstract}
The study of particle correlations is an important instrument to understand the nature of relativistic heavy ion collisions. Using a wealth of new data available from the recent heavy ion runs of Large Hadron Collider at CERN it becomes possible to study particle correlations in different collisions systems under the same conditions. The results of several recent measurement performed by the ATLAS experiment are reviewed in this proceeding. Measurements are
performed using various techniques in \pp, \pPb and \PbPb collisions at the energies \sqs, \sqn from 2.76 to 13~TeV. The results are compared between the systems having the same charged particle multiplicities in the final state, but different initial geometries. Results for multiplicity correlations, two-particle and muti-particle correlations measured in different techniques are presented and discussed. The goal of these complementary analyses is to further understanding the nature of fluctuations observed in small collision systems.
\end{abstract}

\begin{keyword}
heavy ion collisions \sep multi-particle correlations


\end{keyword}

\end{frontmatter}


\section{Introduction}
\label{sec:intro}

Space time evolution of the hot and dense medium created in heavy-ion collisions at high energy can be studied by measuring correlations between particles. The correlations arise as a result of strongly fluctuations geometry in the initial stage of the collision on an event-by-event basis, which are being transformed into correlations between particles in the final state. The underlying physics of this process can be described by relativistic viscous hydrodynamics~\cite{Gale:2013da,Heinz:2013th}. Studies of multi-particle correlation in the transverse to the heavy-ion collision plane revealed strong modulation of the particle emission azimuthal angle, commonly referred to as the anisotropic flow. The measurements of flow coefficients $v_{n}$~\cite{Adare:2011tg,ALICE:2011ab,Chatrchyan:2013kba,Aad:2013xma} and their event-by-event fluctuations~\cite{Aad:2014fla,Aad:2015lwa,Adam:2015eta} put constraints on the properties of the medium.

Multi-particle correlations in the transverse plane and two-particle correlations in particular, have also been studied in small systems like \pp~\cite{Khachatryan:2010gv,Aad:2015gqa,Khachatryan:2015lva} and \pPb~\cite{CMS:2012qk,Abelev:2012ola,Aad:2012gla,Chatrchyan:2013nka,Aad:2014lta} collisions, and these studies have revealed features that bear considerable similarity to those observed in heavy-ion collisions. These findings, typically conducted in high multiplicity events generated many theoretical interpretations~\cite{Dusling:2015gta}, and much discussion as to whether the mechanisms that result in the observed correlations are the same in different collision systems. These findings bring to focus the necessity of adequate comparison between collision systems of different geometry obtained under similar  conditions. In particular, they suggest further investigation of the \pp system that by itself contains volume for a variety of final stages that can also be a manifestation of initial stage fluctuations and hydro-like system evolution.

\section{Forward-backward correlations}
\label{sec:fb}

\begin{figure*}[htb!]
\centering
\includegraphics[width=0.99 \textwidth]{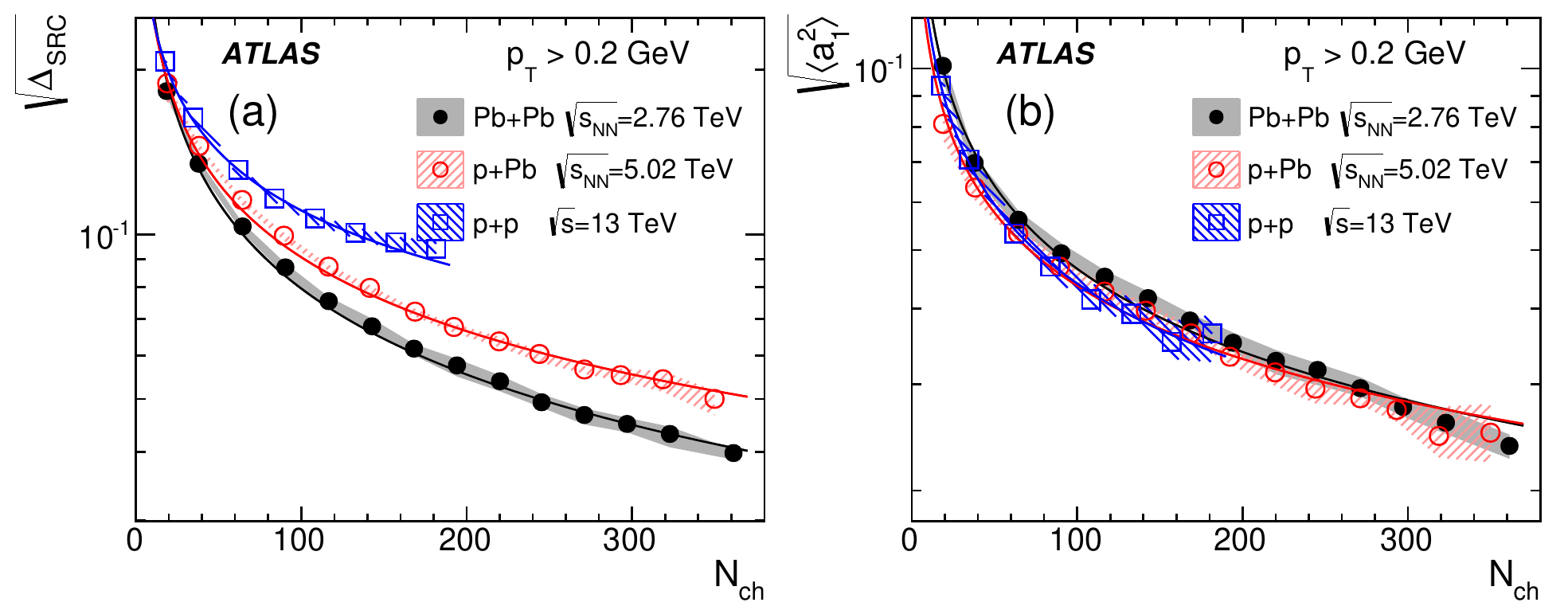}
\caption{The estimated magnitude of the short-range component $\sqrt{\Delta_{\rm{SRC}}}$ (left panel) and $\sqrt{\langle a_{1}^{2}\rangle}$ (right panel) values as a function of $N_{\rm{ch}}$ for all-charge pairs in \PbPb (solid circles), \pPb (open circles), and \pp (open squares)
collisions. The shaded bands represent the systematic uncertainties, and the statistical uncertainties are smaller than the symbols. From publicaiton~\cite{Aaboud:2016jnr}.}
\label{fig:src_lrc}
\end{figure*}

The ATLAS experiment~\cite{PERF-2007-01} measured the two-particle pseudorapidity correlations in \sqn = 2.76~TeV \PbPb, \sqn = 5.02~TeV \pPb, and \sqs = 13 TeV \pp collisions~\cite{Aaboud:2016jnr}. The analysis uses tracks reconstructed with $|\eta| < 2.4$ and transverse momentum $\pT > 0.2$~GeV. The pairs formed with the tracks show patterns that can be characterized as long-range multiplicity correlations (LRC), and short-range correlations (SRC). Their separation exploits the fact that the SRC is observed to be much stronger for opposite-charge pairs than for the same-charge pairs, while the LRC is found to be similar for the two charge combinations. The magnitudes of the SRC in \pPb is found to be larger in the proton-going direction than the lead-going direction, reflecting the fact that the particle multiplicity is smaller in the proton-going direction. This is consistent with the observation that the SRC strength increases for smaller multiplicity. The strength of the SRC correlation plotted as a function multiplicity is shown in the left panel of Fig.~\ref{fig:src_lrc} taken from Ref.~\cite{Aaboud:2016jnr}. The distributions shown in the plot can be approximated by the power law with very different indices in the three plotted systems.  In contrast, the first Legendre polynome coefficient in the expansion of the LRC function that is shown in the right panel of Fig.~\ref{fig:src_lrc}, exhibits similar, power law behavior as the SRC, but with the same index values in all three measured collision systems.

\section{Long-range two-particle azimuthal anisotropies}
\label{sec:template}
\begin{figure}[htb!]
\centering
\includegraphics[width=0.49 \textwidth]{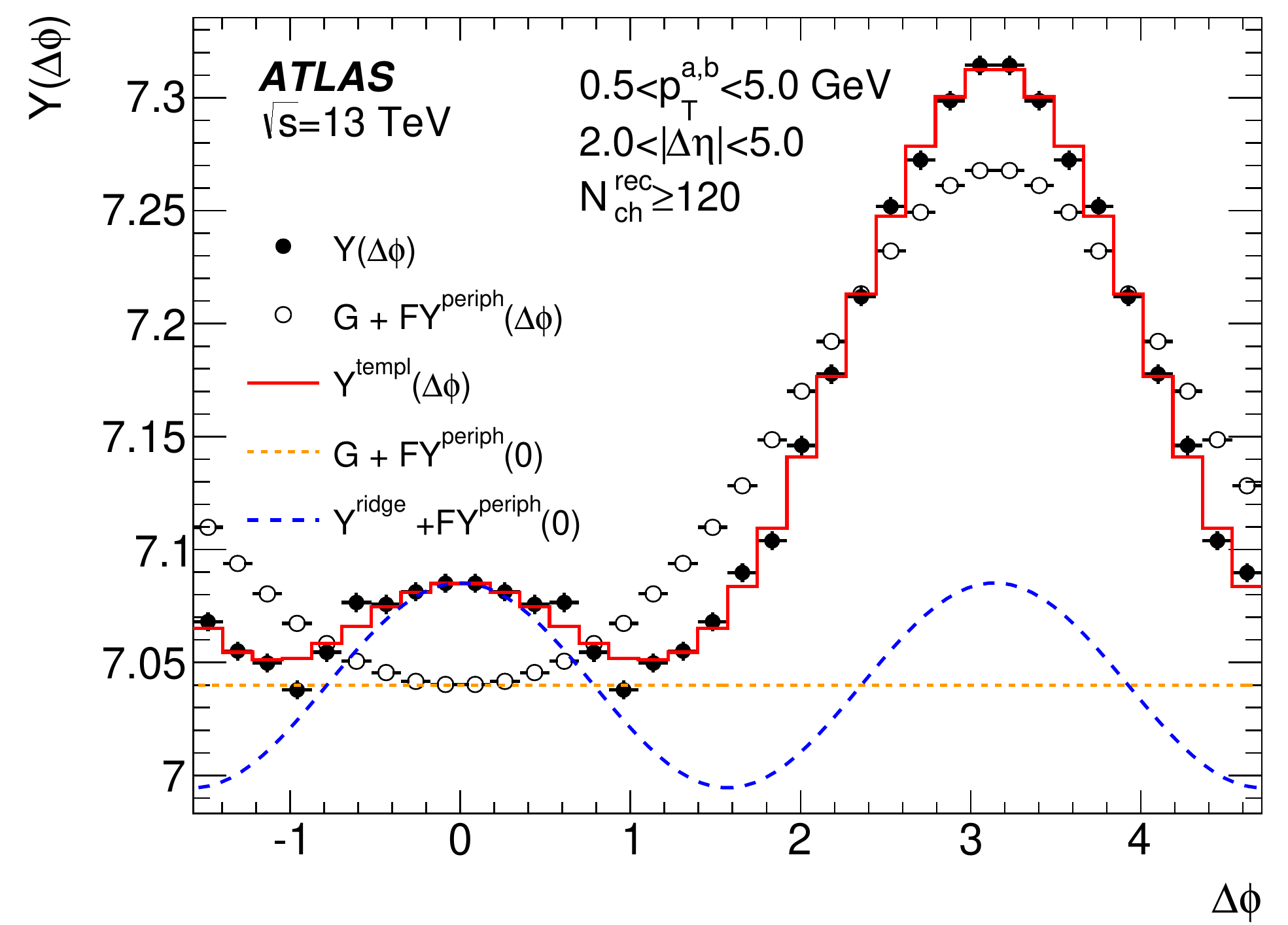}
\caption{Template fit to the particle yield in 13 TeV \pp collisions for charged particle pairs with $0.5 < p_{\rm{T}}^{a,b} < 5$ GeV and $2 < |\Delta\eta| < 5$. The plot corresponds to more than 120 reconstructed tracks. From publication~\cite{Aad:2015gqa}.}
\label{fig:template}
\end{figure}

ATLAS measured~\cite{Aad:2015gqa} the two-charged-particle correlations in \sqs=13~and~2.76 TeV \pp collisions using a new template fitting procedure. In this novel approach, it is shown that the per-trigger-particle yields for $|\Delta\eta|>2$ are described by a superposition of the yields measured in a low-multiplicity interval and a constant modulated by $cos(2\Delta\phi)$ as shown in Fig.~\ref{fig:template}.

\begin{figure*}[h!]
\centering
\includegraphics[width=0.49 \textwidth]{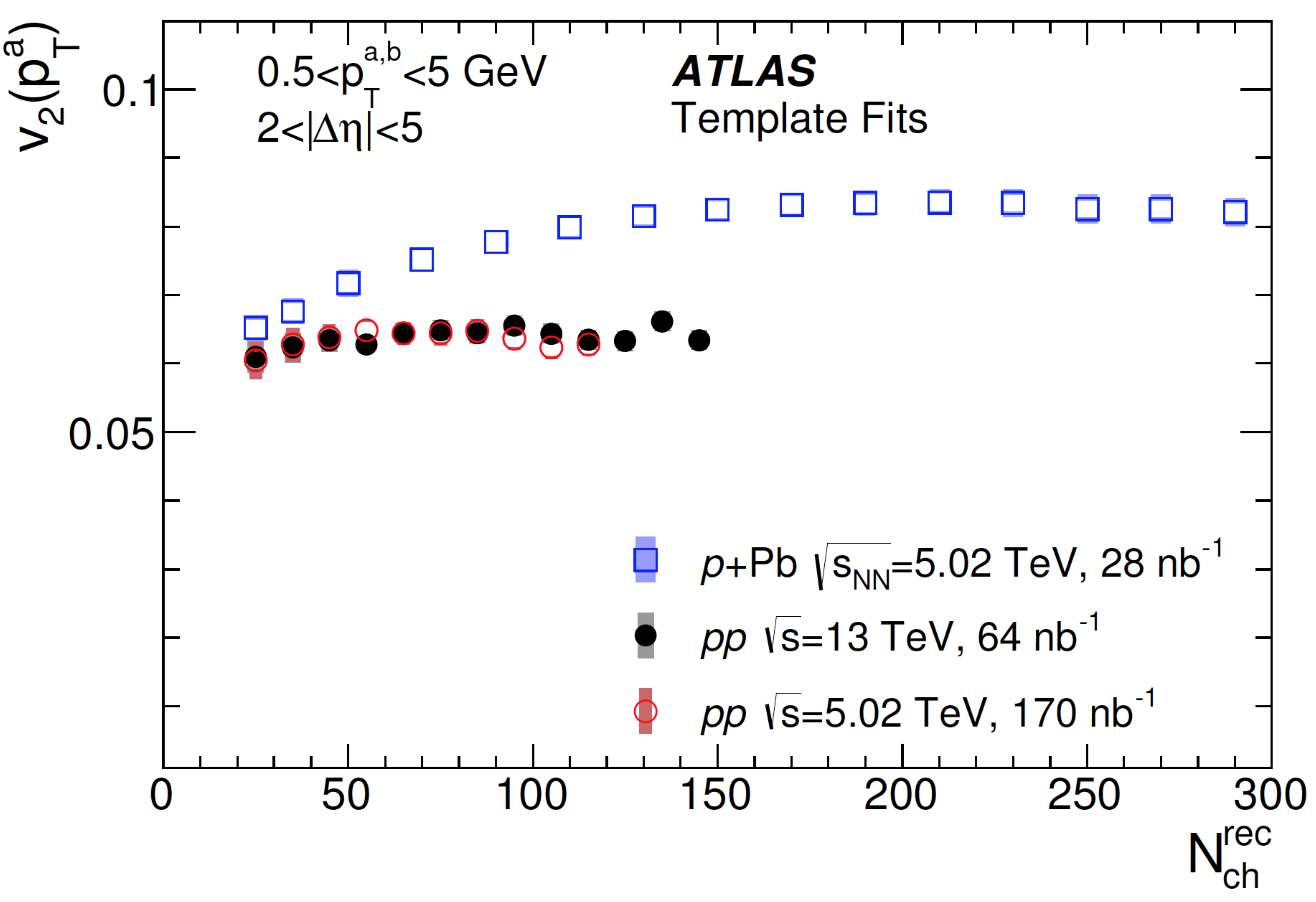}
\includegraphics[width=0.49 \textwidth]{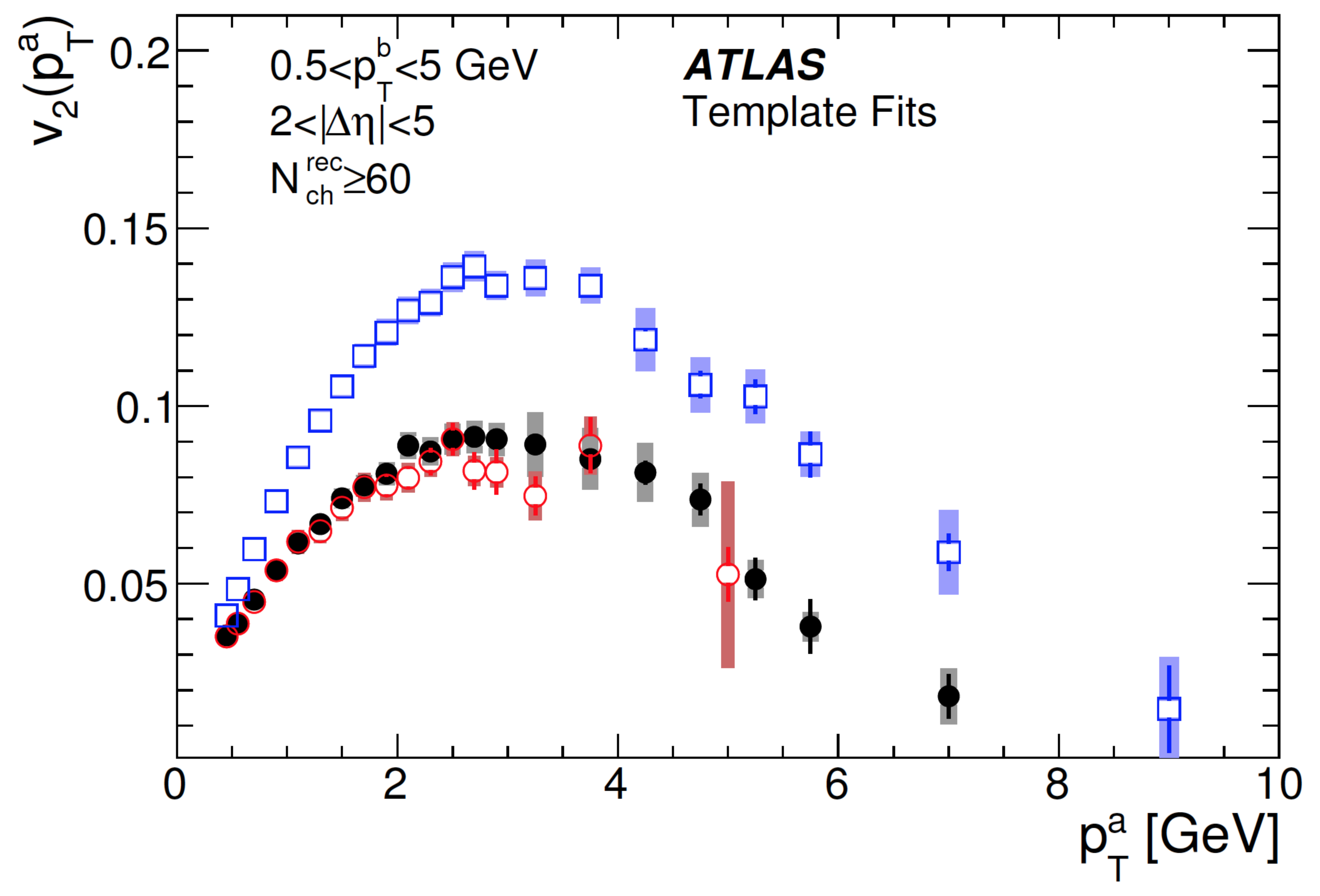}
\caption{Left: the $v_{n}$ obtained from the template fitting procedure in the 13~TeV \pp, 5.02~TeV \pp, and 5.02~TeV \pPb data, as a function of multiplicity. The results are for $0.5< p_{\rm{T}}^{a,b} < 5$~GeV. Right: the \pT dependence of the $v_{n}$ for multiplicity range above 60 tracks.  The error bars and shaded bands indicate statistical and systematic uncertainties, respectively. From publication~\cite{Aaboud:2016yar}.}
\label{fig:template_results}
\end{figure*}

Unlike previous two-particle correlations analyses relying on the "zero yield at minimum" hypothesis to separate the ridge from the peak at $|\Delta\phi|\approx\pi$ and therefore has limited applicability in events with low number of tracks, this template method explicitly accounts for the shape of correlations in peripheral events. This shape is shown with open circles in Fig.~\ref{fig:template} including a constant combinatorial contribution. This shape is fit to the data points (closed circles) to derive the ridge component, shown in the figure with a dashed line.

Using the template procedure, the correlation functions measured at two energies in \pp collisions show a ridge whose strength increases with multiplicity. 
The extracted Fourier coefficients, exhibit factorization, which is characteristic of a global modulation of the per-event single-particle distributions also seen in \pPb and \PbPb collisions. The amplitudes of the single-particle modulation, are found to be independent of multiplicity and agree between 2.76 and 13~TeV within uncertainties. The amplitudes rise with \pT until 3 GeV, and then decrease with higher \pT, following a trend similar to that observed in \pPb and \PbPb. These results suggest that the ridges in \pp and \pPb collisions may arise from a similar physical mechanism which does not have a strong \sqs dependence.

The template method used in~\cite{Aad:2015gqa} is used to measure higher order harmonics in \pp collisions at $\sqs=5.02$~and~13~TeV and also to the \pPb collisions~\cite{Aaboud:2016yar}. The results shown in Fig.~\ref{fig:template_results} demonstrate that the \pPb $v_{2}$ increases with multiplicity while the \pp $v_{2}$ does not. Results published in~\cite{Aaboud:2016yar} also show that the same trend is seen for the $v_{3}$, which  has a larger difference between \pPb and \pp, compared to $v_{2}$, but it also has larger systematics uncertainties. Results for $v_{4}$ are consistent between measured systems and only weakly depend on multiplicity.

As a function of \pT, the $v_{2}$ results shown in the right panel of Fig.~\ref{fig:template_results} for \pp and \pPb systems display similar trends. 
The $v_{2}$ values for the 5.02~and~13~TeV \pp data agree within uncertainties but are lower than for \pPb. The \pT dependence of the higher harmonics is similar to that of $v_{2}$ at low \pT, whereas the \pPb results increase more rapidly.

\section{Multi-particle azimuthal correlations}
\label{sec:cumulants}

\begin{figure*}[ht!]
\centering
\includegraphics[width=0.49 \textwidth]{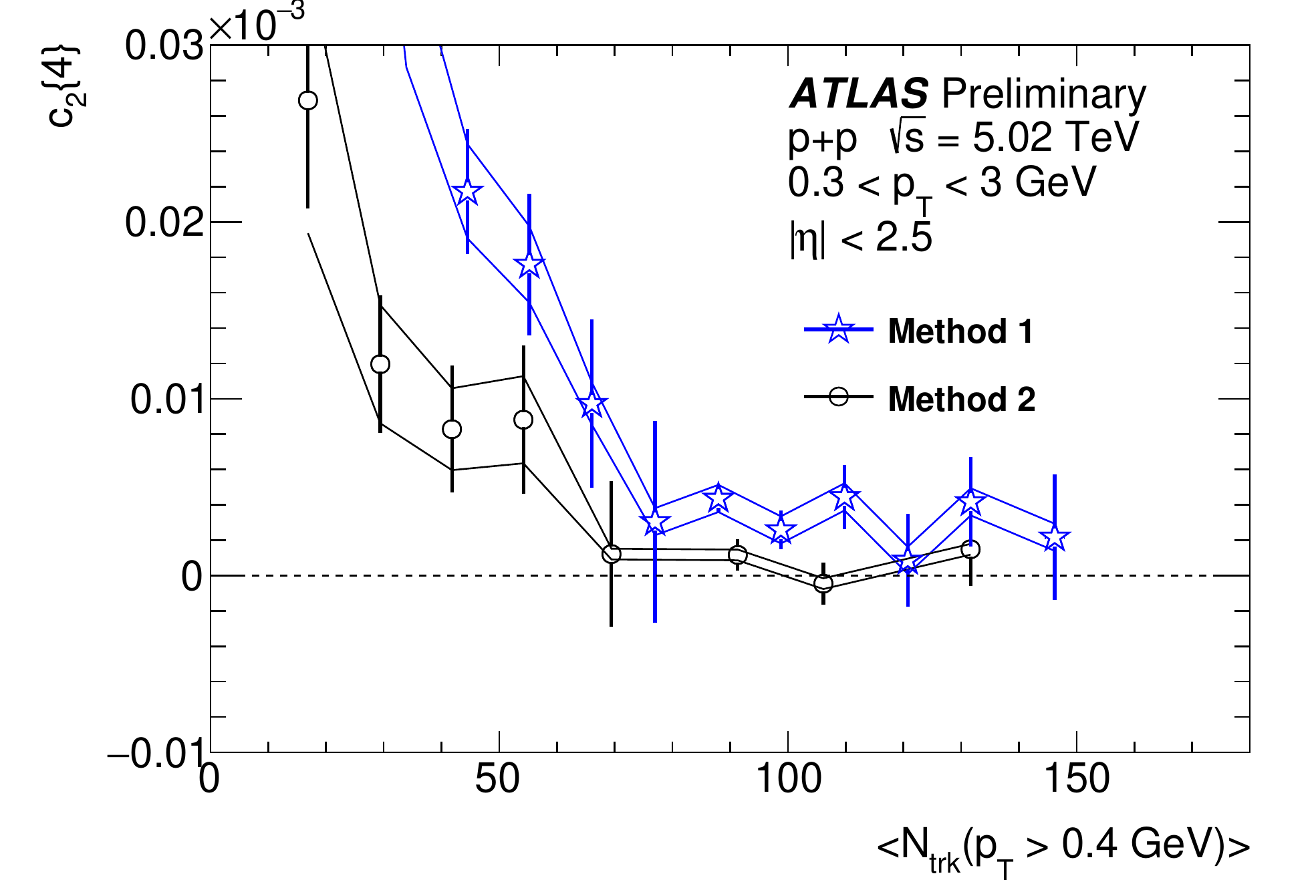}
\includegraphics[width=0.49 \textwidth]{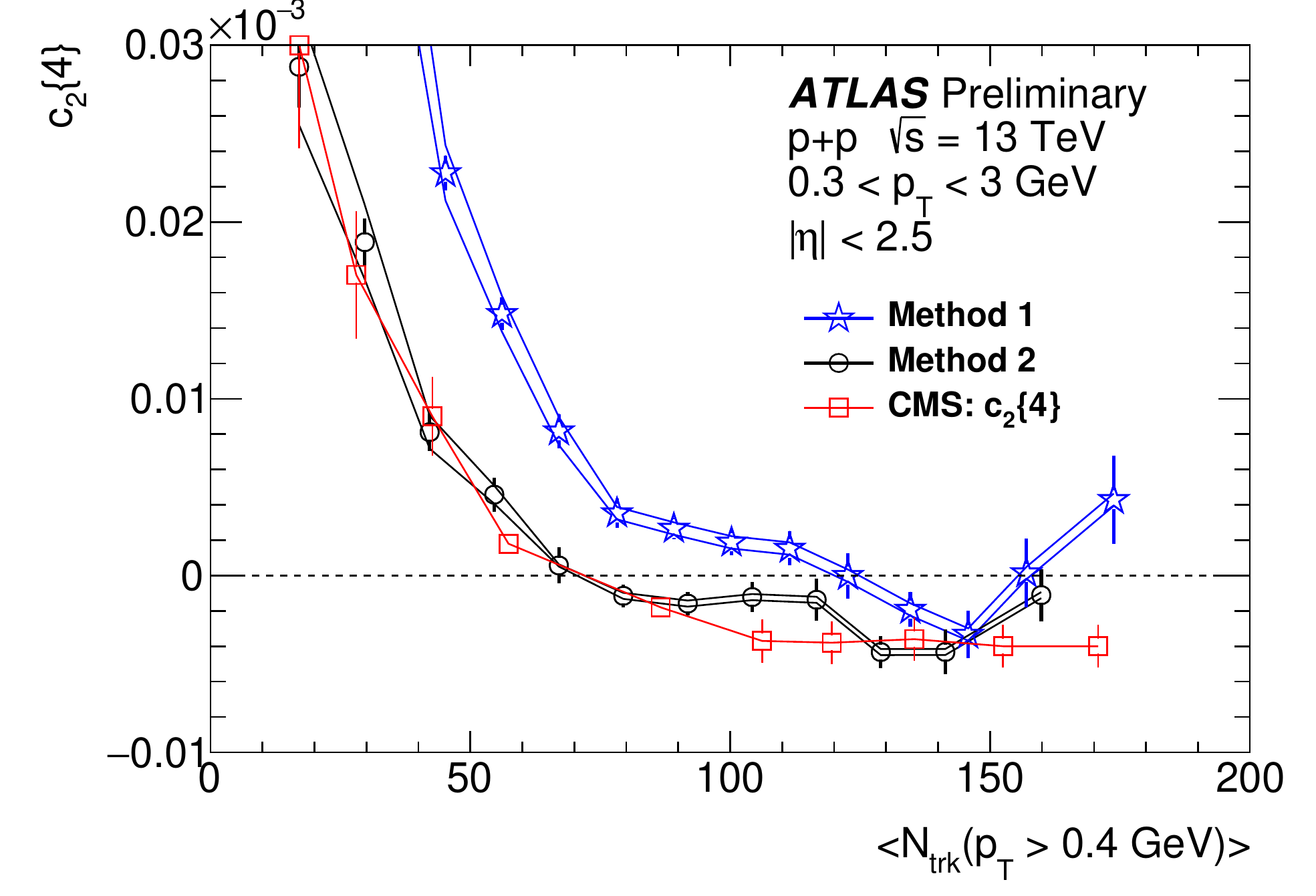}
\caption{The comparison of cumulants obtained with two different methods for \pp collisions at 5.02~TeV (left) and 13~TeV (right). Plots show results for reference particles with $0.3<\pT<3.0$~GeV. The error bars and shaded bands denote statistical and systematic uncertainties, respectively. The CMS results from~\cite{Khachatryan:2016txc} are also shown as open squares. From publication~\cite{ATLAS-CONF-2016-106}.}
\label{fig:cumulants}
\end{figure*}

Four-particle cumulants are measured by ATLAS in \pp collisions at $\sqs = 5.02$~and~13~TeV~\cite{ATLAS-CONF-2016-106}. Two analysis methods used in this measurement are different in their sensitivity to correlations not related to the initial collision geometry (referred to as non-flow correlations), for example from resonance decays, jet production, quantum interference or energy-momentum conservation. Implementation of the methods differs in calculating cumulants in fixed bins of all tracks in the first method or of reference tracks in the second. Bothline methods are used to calculate $c_{2}\{4\}$ cumulant. It is found in the data, and also in events generated by PYTHIA 8, that multiplicity fluctuations give a negative contribution to $c_{2}\{4\}$ that is characteristic of the collective-like effects. Figure~\ref{fig:cumulants} compares results of the two methods used to measure cumulants in 5.02~TeV \pp data shown on the left and in 13~TeV data on the right. Method~2, which is more susceptible to non-flow correlations, yields smaller values of $c_{2}\{4\}$, eventually becoming negative, which formally allows extracting the flow-like $v_{2}\{4\}$ for multiplicity above 80 tracks in 13~TeV data. The results of this method are consistent with the recent results from CMS~\cite{Khachatryan:2016txc}, obtained with the analysis method comparable to the Method 2.

The measurements of cumulants with Method 1 does not provide evidence for collectivity. However, even though free of multiplicity fluctuations, the standard procedure used to calculate $c_{2}\{4\}$ can be still biased by non-flow correlations, and so the question of collectivity from a cumulant-based approach remains open.

\section{Aknowledgements}
\label{sec:akn}
This research is supported by the Israel Science Foundation (grant 1065/15) and by the MINERVA Stiftung with the funds from the BMBF of the Federal Republic of Germany.



\nocite{*}
\bibliographystyle{elsarticle-num}
\bibliography{milov_sasha_parallel_VI_IS}







\end{document}